\documentclass[useAMS,usenatbib]{mn2e}
\usepackage[pdfpagemode={UseOutlines},bookmarks,bookmarksopen,colorlinks,citecolor={blue},urlcolor={red}]{hyperref}
\usepackage[flushleft]{threeparttable}
\usepackage{booktabs,caption}
\usepackage{graphicx}
\usepackage{amsmath}
\usepackage{subfigure}
\usepackage{amssymb}
\usepackage{tabularx}
\usepackage{natbib}
\usepackage{float}
\usepackage{etoolbox}

\usepackage{color}

\usepackage{natbib}
\usepackage[usenames,dvipsnames]{xcolor}

\usepackage[normalem]{ulem}

\newcommand{\bea}{\begin{eqnarray}}
\newcommand{\eea}{\end{eqnarray}}

\makeatletter

\patchcmd{\NAT@citex}
  {\@citea\NAT@hyper@{%
     \NAT@nmfmt{\NAT@nm}%
     \hyper@natlinkbreak{\NAT@aysep\NAT@spacechar}{\@citeb\@extra@b@citeb}%
     \NAT@date}}
  {\@citea\NAT@nmfmt{\NAT@nm}%
   \NAT@aysep\NAT@spacechar\NAT@hyper@{\NAT@date}}{}{}

\patchcmd{\NAT@citex}
  {\@citea\NAT@hyper@{%
     \NAT@nmfmt{\NAT@nm}%
     \hyper@natlinkbreak{\NAT@spacechar\NAT@@open\if*#1*\else#1\NAT@spacechar\fi}%
       {\@citeb\@extra@b@citeb}%
     \NAT@date}}
  {\@citea\NAT@nmfmt{\NAT@nm}%
   \NAT@spacechar\NAT@@open\if*#1*\else#1\NAT@spacechar\fi\NAT@hyper@{\NAT@date}}
  {}{}
  
  \makeatother

\newcommand{\gae}{\lower 2pt \hbox{$\, \buildrel {\scriptstyle >}\over {\scriptstyle \sim}\,$}} 

\newcommand{\lae}{\lower 2pt \hbox{$\, \buildrel {\scriptstyle <}\over {\scriptstyle \sim}\,$}} 

\title[Lack of NCs in dSphs]{Lack of nuclear clusters in dwarf spheroidal galaxies: implications for massive black holes formation and the cusp/core problem}

\author[Arca-Sedda, M. and Capuzzo-Dolcetta, R.]{Manuel Arca-Sedda$^{1}$\thanks{m.arcasedda@gmail.com} Roberto Capuzzo-Dolcetta$^{1}$\\
Dept. of Physics, Sapienza, University of Rome, Piazzale Aldo Moro 5, I-00185, Rome (Italy)}

\begin{document}
\date{Revised to 09/2016}

\pagerange{\pageref{firstpage}--\pageref{lastpage}} \pubyear{2015}

\maketitle

\label{firstpage}

\maketitle

\begin{abstract}
One of the leading scenarios for the formation of nuclear star clusters in galaxies 
is related to the orbital decay of globular clusters (GCs) and their subsequent merging, though alternative theories are currently debated. The availability of high-quality data for GCs structural and orbital parameters allow to test different nuclear star cluster formation scenarios.
The Fornax dwarf spheroidal (dSph) galaxy is the heaviest satellite of the Milky Way and it is the only known dwarf spheroidal hosting 5 GCs, whereas there are no clear signatures for the presence of a central massive black hole. For this reason, it represents a suited place to study the orbital decay process in dwarf galaxies. In this paper we model the future evolution of the Fornax GCs simulating them and the  host galaxy by means of direct $N$-body simulations. Our simulations take in account also the gravitational field generated by the Milky Way.
We found that if the Fornax galaxy is embedded in a standard Cold Dark Matter Halo, the nuclear cluster formation would be significantly hampered by the high central galactic mass density. In this context, we discuss the possibility that infalling GCs drive the flattening of the galactic density profile, giving a possible alternative explanation to the so-called cusp/core problem. Moreover, we briefly discuss the link between GC infall process and the absence of massive black holes in the centre of dSphs.
\end{abstract}


\section{Introduction}

Among all the dwarf spheroidal (dSph) satellite galaxies of the Milky Way (MW), Fornax is the only one hosting five globular clusters (GCs), at projected distances from its centre smaller than 1 kpc. On the other hand, this galaxy does not exhibit any evidence of a clear central over density or the presence of a massive black hole (MBH). Several works have pointed out that the Fornax dSph lies inside a huge dark matter halo (DMH) \citep{mateo}. The presence of GCs, named Fornax 1, 2, 3, 4 and 5, represents an open puzzle since the action of dynamical friction (df) should have dragged them into the galactic centre a few Gyrs ago. 
Indeed, these old, metal-poor GCs have ages exceeding $10$ Gyr \citep{buonanno98,buonanno,lrs}, a value significantly larger than their df decay time \citep{tremaine,ohlirich}. However, it has been widely demonstrated that the standard approach for studying the df process, developed so far by \cite{Cha43I}, poorly describes the motion in axisymmetric and triaxial \citep{OBS,Pes92} or spherical and cuspy galaxies \citep{AntMer12,ascd14df,petts15,petts16}. Moreover, since the orbital energy loss depends primarily on the host galaxy structure, a reliable description of the GCs motion requires a detailed knowledge of the Fornax mass distribution.
Though the Fornax dSph is believed, like all the dSphs, to be enclosed within a DMH, its density profile seems to differ significantly from the standard prediction of the $\Lambda$-CDM paradigm, $\rho(r) \propto r^{-1}$ \citep{walker,NFW96}. Indeed, observational constraints suggest a flatter density profile \citep{flores,moore,gilmore,cowsik,jardel}.
Hence, the study of the dynamical evolution of Fornax GCs can shed light on the structure of dSph galaxies, and, in particular, on their DM content.

Currently, several scenarios have been proposed and debated to explain the dynamical evolution of the Fornax and its GCs.

For instance, some authors proposed that this galaxy is characterised by a flat density profile, with a core extending up to $300$ pc \citep{strigari}, while others argued that it is the product of a merger event \citep{olsew,coleman04,yozin}. 
\cite{angus} proposed, instead, that the assumption of modified Newtonian dynamics (MOND) \citep{milgrom83} leads to a direct interpretation of the observed dynamical status of the Fornax GCs. 
Another possibility relies upon that df halts its role within the region where the density profile flattens \citep{Goerdt,Read,Gual08,ascd14df}.

Using recent observational data, in a companion paper we revisited the so called ``timing'' problem, showing that all the GCs have been formed within the Fornax tidal radius, even in the case of a standard CDM density profile for the Fornax halo \citep{ascd15frn}.

Intriguingly, the presence of 5 GCs moving in the inner region of the galaxy coincides with the absence of a bright nucleus.
Bright nuclei, often referred to as nuclear clusters (NCs), are very dense stellar systems with masses in the range $10^{8}-10^{11}$ M$_\odot$ observed in galaxies of all the Hubble types \citep{vanberg,cote04,cote06,turner}. Some authors proposed an ``in-situ'' origin for NCs \citep{McLgh,nayakshin,aharon15}; another possibility is that GCs decay toward the centre of their host galaxy, where they collide and merge, driving the formation of a NC and, even, the accretion of a black hole seed therein \citep{TrOsSp,Dolc93}. This channel of formation, named ``dry-merger'' scenario, has been successfully investigated both on a theoretical and numerical side \citep{DoMioB,AMBANT,Antonini13,ascd14nsc}, and explains satisfactorily the observed correlations among NCs properties and those of their host galaxies \citep{ascd14df,LGH,scot}. 
Recently, some authors have shown that the tidal forces induced by a very massive BH can tidally disrupt the infalling GCs, thus suppressing the formation of a NC \citep{bekkiGr, Antonini13, ascd14nsc, ascd15he, ascd15lack}.

In a recent paper, \cite{ascd15he} investigated the future evolution of the 11 young massive clusters located in the dwarf starburst galaxy Henize 2-10. This work is the first showing that the formation of a NC is a process that can occur in a galaxy containing a pre-existing MBH, even if the tidal field induced by the massive BH and by the background stars can alter significantly the NC formation process. In the same framework, in this paper we simulate the future evolution of the Fornax globular cluster system (GCS) making use of direct $N$-body modelling. 
The availability of high-quality data on the GCs structural and orbital parameters and on the host galaxy would allow us to investigate whether their future evolution may lead to the formation of a nucleated region. This has a double aim: i) study the ongoing orbital decay process and put constraints on the possible formation of nuclear clusters in dwarf spheroidal galaxies, and, ii) extrapolate our results backward in time in order to obtain hints about the GCs origin and past evolution.

The paper is organized as follows: in Section \ref{model} we describe the matter distribution adopted to model the galaxy and the GCs, and the numerical method used; in Section \ref{results} we present the results of the $N$-body simulations and discuss them in Section \ref{discussion}; finally, in Section \ref{conclusion} we draw the conclusions of this work.

\section{Modelling the Milky Way, the Fornax dSph and its globular clusters}
\label{model}

In this work we present results from a series of direct $N$-body simulations in which we modelled the orbital decay of the 5 GCs orbiting in the Fornax dSph galaxy. To carry out the simulations, we used the direct $N$-body code \texttt{HiGPUs} \citep{spera}, a 6$^{\rm th}$-order Hermite integrator with block time-step, which runs on graphic processing units (GPUs), fully exploiting the advantages of parallel machines. 

In all the simulations performed, the MW is modelled as an external, analytical, field, whereas the Fornax dSph and all the GCs have been represented self-consistently as $N$ ``particles''. The number of particles that should be used for modelling the whole dSph exceeds $10^8$, which is too much to allow a reasonable computational time. 
To overcome this problem, we: i) limited the representation of the galaxy to an inner region of $\sim 2$ kpc, as explained more in detail below, and, ii), we allowed a difference between the individual mass of the particles in the dSph, $m_F$, and in the GCs, $m_{\rm GC}$, such that $m_F/m_{\rm GC}=10$.

This implies a number of particles which is, in the smallest cluster, greater than $3\times 10^3$. The  evaporation time, which is the time-scale over which two-body relaxation is at all important, is $\sim 2$ Gyr for the smallest cluster. Consequently, the evaporation time for all the clusters is significantly larger than the time over which they reach the Fornax centre, thus avoiding possible spurious effects of relaxation due to an exceedingly small number of particles.

\subsection{The Milky Way model}
The MW tidal field was modelled with the \cite{allen} density profile, given by the sum of three different components, bulge, disk and halo. 
The galactic bulge is represented as a Plummer sphere \citep{plum}:
\begin{equation}
\Phi_b(r) = -\frac{M_b}{\sqrt{r^2+a_b^2}}, 
\end{equation} 

the galactic disk is represented by a \cite{MyN} law:

\begin{equation}
\Phi_d(x,y,z) = -\frac{M_d}{\sqrt{x^2+y^2+\left(a_d^2+\sqrt{z^2+b_d^2}\right)^2}},
\end{equation} 

and the halo is described by a \cite{NFW96} profile:
\begin{equation}
\Phi_h(r) = -\frac{M_h(r/a_h)^{2.02}}{r(1+(r/a_h)^{1.02})}-\frac{M_h}{1.02a_h}\left[F(r)-F(b_h))\right],
\end{equation}

where $F(r)$ is defined as:

\begin{equation}
F(r) = -\frac{1.02}{1+(r/a_h)^{1.02}}+\ln\left[1+(r/a_h)^{1.02}\right].
\end{equation}

In Eq.s (1--4), $a_b$, $a_d$, $b_d$, $a_h$ are scale lengths, while $M_b$, $M_d$ and $M_h$ are masses, 
all listed in Table \ref{astab}. We notice here that we used the same parameters adopted in \cite{ascd15frn}.

\begin{table}
\caption{}
\centering{Main parameters of the Milky Way profile}
\begin{center}
\begin{tabular}{cccc}
\hline
\hline
    & $M$ & $a$ & $b$ \\
	& $(10^{10}$ M$_\odot)$ & (kpc) & (kpc) \\
\hline
Bulge & 1.4 & 0.39 & - \\
Disk  & 8.6 & 5.32 & 0.25 \\
Halo  &11 & 12 & 100 \\
\hline
\end{tabular}
\end{center}
\begin{tablenotes}
\item Column 1: component of the MW. Column 2: mass scale of the component. Column 3-4: length scales of the component.
\end{tablenotes}
\label{astab}
\end{table}

In order to describe properly the motion of Fornax around the Milky Way, we selected the following initial conditions (ICs) in the MW reference frame:
\begin{align}
&{\bf R}_{F0}({\rm kpc}) = (0,66,220),&\\
&{\bf V}_{F0}({\rm kms}^{-1}) = (-206,0,193).&
\label{inposvel}
\end{align}
This choice implies a pericenter and velocity in agreement with observational estimates \citep{buonanno, dinescu}.

Setting the MW tidal field allows us to determine the Fornax tidal radius, $R_t$, along its trajectory, which represents an indication of the dSph boundary.
A good approximation of this length is given by:
\begin{equation}
R_t = \left(\frac{GM_F}{\omega^2+({\rm d^2\Phi_{MW}}/{\rm dr^2})_{R_{\rm F_p}}}\right)^{1/3},
\label{tidR}
\end{equation}

where $M_F$ is the Fornax mass, $\omega$ its angular velocity (assumed constant) and $R_{F_p}$ the Fornax pericenter distance. Assuming $M_F = 10^8$ M$_\odot$ and $R_{\rm F_p} = 138$ kpc \citep{buonanno}, we obtained $R_t = 5$ kpc.

\subsection{The Fornax dSph model}
According to observational data, a suitable expression for the Fornax mass density profile is a slightly modified Dehnen's law \citep{Deh93}:
\begin{equation}
\rho(r)=\frac{(3-\gamma)M_F}{4\pi r_F^3\cosh{\left(r/r_{\rm cut}\right)}}\left(\frac{r}{r_F}\right)^{-\gamma}\left(\frac{r}{r_F}+1\right)^{-4+\gamma},
\label{rhotr}
\end{equation}
where, again, $M_F$ is the Fornax mass, $r_F$ its scale radius, and $\gamma$ defines the inner slope of the density profile. The modification comes from the hyperbolic cosine at denominator, which allows an exterior cut in the sampling by particles, whose length scale is given by $r_{cut}$. 

We developed two models with characteristics in agreement with observations, assuming (model D05) $M_F = 1.48 \times 10^8$ M$_\odot$, $r_F=0.301$ kpc and $\gamma = 0.5$, and (model D1) $M_F = 1.48 \times 10^8$ M$_\odot$, $r_F=0.391$ kpc and $\gamma = 1$.

\begin{figure}
\centering
\includegraphics[width=8cm]{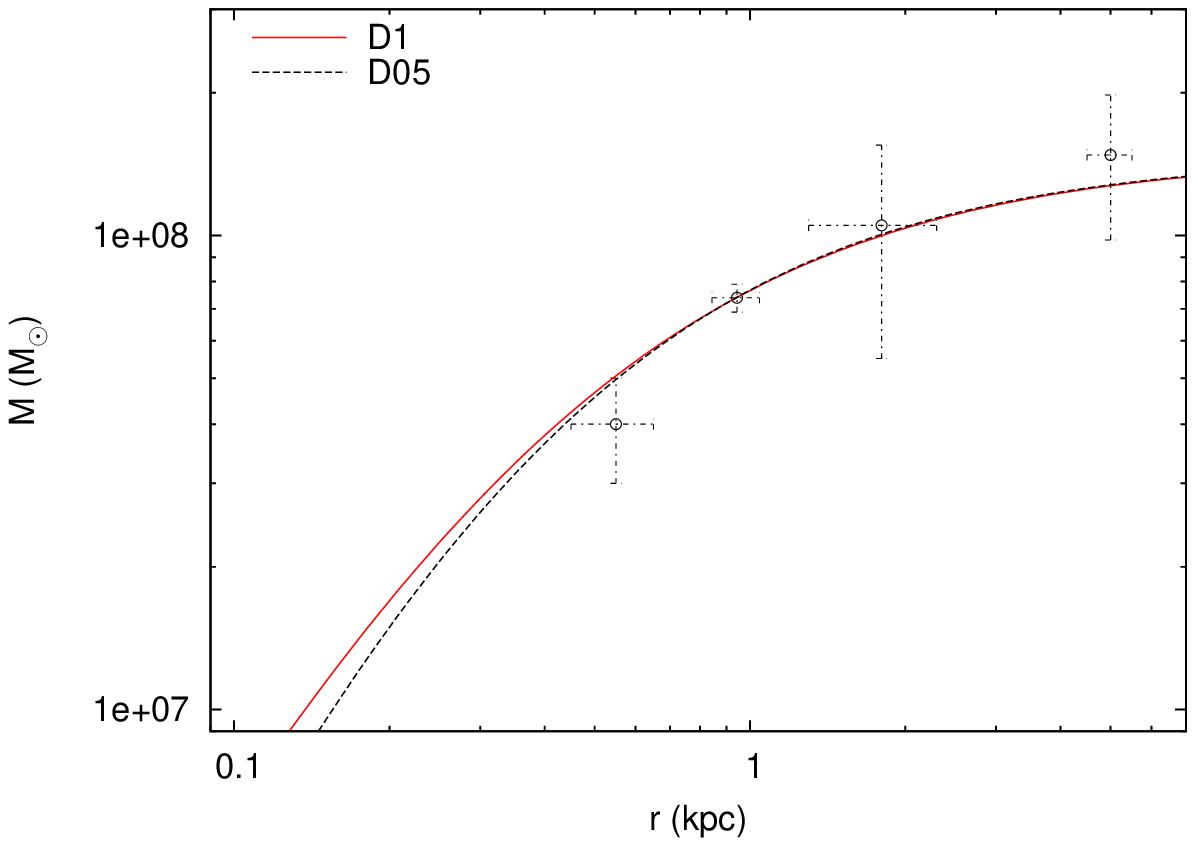}
\caption{Cumulative mass profile for model D05 (straight red line) and model D1 (dotted black line) compared with observational data provided by \citet{walker},  \citet{wolf}, \citet{Walkr} and \citet{cole}.}
\label{masspro}
\end{figure}

These choices lead to models characterised by a radial mass profile in good agreement with observed mass estimates based on kinematical data, as shown in Figure \ref{masspro}.
It is worth noting that the two profiles are almost indistinguishable for $r>1$ kpc. 
Table \ref{Frnmod} summarizes the parameter choices we made for both models D05 and D1.

\begin{table*}
\caption{}
\centering{Main parameters of the Fornax dSph model}
\begin{center}
\begin{tabular}{lcccccc}
\hline
\hline
  ID  & $M_F$ & $r_F$ & $\gamma$ & $r_{\rm cut}$ & $M_{\rm cut}$ & N\\
	& $(10^{8}$ M$_\odot)$ & (kpc) & & (kpc) & $(10^{8}$ M$_\odot)$ &  \\
\hline
D05 & 1.48 & 0.301 & 0.5 & 2 & 1.06 & 967889\\
D1  & 1.48 & 0.391 & 1.0 & 2 & 1.06 & 967188\\
\hline
\end{tabular}
\end{center}
\begin{tablenotes}
\item Column 1: model name. Column 2: total mass of the model. Column 3: model scale radius. Column 4: slope of the density profile. Column 5: truncation radius. Column 6: mass enclosed within $r_{\rm cut}$. Column 7: number of particles used to model the system.
\end{tablenotes}
\label{Frnmod}
\end{table*}

\subsection{The GCs models}

To model the Fornax GCs, we used data collected in several papers \citep{hodge61,rodgers94,mackey,cole}, as summarized in Table \ref{GCobs}.

The initial conditions for GC velocities are picked from the energy dependent distribution function that gives the background density  of Eq. \ref{rhotr}. Under this requirement, we added the constraint that the GC positions in the x-y plane are the same as observed, and that the z component of the velocity is equal to the observed line-of-sight velocity. 
Moreover, we required that the GC 3D spatial positions are close to their observed projected positions. 

While the other choices are dictated by observational constraints, the latter assumption is related to the so called Fornax \textit{timing problem}. The problem resides in that the Fornax GC ages are significantly greater than their estimated dynamical friction decay times, which makes highly unlikely their present observed positions in the galaxy. As pointed out above, one of the solutions proposed to solve such a puzzle is the claim that GC dynamical friction decay lead them to stall many Gyr ago in the case of a Fornax cored density profile \citep{read06,cole}. Indeed, as shown by \cite{Gual08}, df stalls in cored systems when the satellite mass equals the galactic mass enclosed within its orbits. This holds also for cusped density profiles, although with some quantitative differences \citep{ascd14df}. 

Assuming, thus, that GC infall ceased many Gyr ago, in order for GCs to survive they needed to resist again tidal shattering for, at least, $20$ orbital revolution around the Fornax nucleus. This because the GC ages are all around $\sim 10$ Gyr and the Fornax crossing time is, roughly, $\sim \sqrt{R_t^3/M_F} = 0.4$ Gyr. A strong tidal field would have, on the other side, likely shattered the clusters after just few orbital periods \citep{ascd15lack}. 
Therefore, the choice of setting $z=0$, which maximizes both tidal forces and df effects, allows us a cleaner view of how these two processes affect the GCs dynamics.
In addition, it is worth noting that the likelihood that GCs spatial and projected positions are close each others has been already proposed by several authors to support the possibility that they all formed within the Fornax tidal radius \citep{cole,ascd15frn}.

The internal structure of the GCs is represented by King models \citep{king}, which are fully defined by the GC mass, $M_{\rm GC}$, its adimensional potential well ($W_0$), and the concentration parameter ($c$), defined as the ratio between the GC core radius, $r_{\rm GC,c}$, and its tidal radius, $r_{\rm GC,t}$. To obtain the tidal radius of each cluster, we used the Eq. \ref{tidR} conveniently modified, whereas for $r_{\rm GC,c}$ we assumed the estimates provided by observations \citep{mackey}.
The knowledge of these two quantities allowed us to get $W_0$, since it is directly related to $c$.

Table \ref{GCnbo} lists all the orbital and structural parameters for our models.

\section{Results}
\label{results}

\subsection{The role of dynamical friction}

In the framework of the dry-merging scenario, the main process that drives the formation of a NC is  dynamical friction \citep{TrOsSp,Dolc93}. 
Using a series of highly resolved $N$-body simulations, \cite{ascd14df} developed a fitting formula for the time, $\tau_{\rm df}$, needed to drag a body of mass $M$ toward the centre of its parent galaxy. Such fitting formula has been improved recently by \cite{ascd15he}:

\begin{equation}
\tau_{\rm df} ~({\rm Myr}) = 0.3\sqrt{\frac{r_F^3 ~({\rm kpc}^3)}{M_F ~(10^{11}~M_\odot)}}g(e,\gamma)\left(\frac{M}{M_F}\right)^{-0.67}\left(\frac{r}{r_F}\right)^{1.76},
\label{tdf}
\end{equation}

where $e$ is the orbital eccentricity and 
\begin{equation}
g(e,\gamma) = (2-\gamma) \left[a_1\left(\frac{1}{(2-\gamma)^{a_2}+a_3}\right)(1-e)+e\right],
\end{equation}
with $a_1 = 2.63\pm 0.17$, $a_2 = 2.26\pm 0.08$ and $a_3 = 0.9\pm 0.1$.

Equation \ref{tdf} is particularly well-suited in describing the orbital decay in galaxies characterised by cusped density profiles, though it has been shown that the formula holds also for cored systems.

Equation \ref{tdf}, applied to the GCs of model D05 and D1, gives a preliminary estimate of the time, $\tau_{df}$, they need to plunge in the innermost region of the galaxy. 

Table \ref{GCnbo} lists $\tau_{\rm df}$ for the GCs in models D05 and D1. In most of the cases $\tau_{\rm df}$ exceeds $1$ Gyr.

Moreover, Eq. \ref{tdf} can be used to make a guess the radial positions, $r_0$, of the GCs at their birth.
Indeed, for a given Dehen's galaxy model and for a given value of $e$, the difference between the GC initial decay time and its current value, estimated according to its present position $r$, is an approximate value of its age, $\tau$. 
This means $= \tau = \tau_{\rm df}(r_0)-\tau_{\rm df}(r)$,
which, solved for $r_0$ gives an estimate of the GC birthplace galactocentric distance:
\begin{equation}
r_0 = r \left(1+\frac{\tau}{\tau_{\rm df}(r)}\right)^{0.57}.
\label{IC0}
\end{equation}

\begin{figure}
\includegraphics[width=8cm]{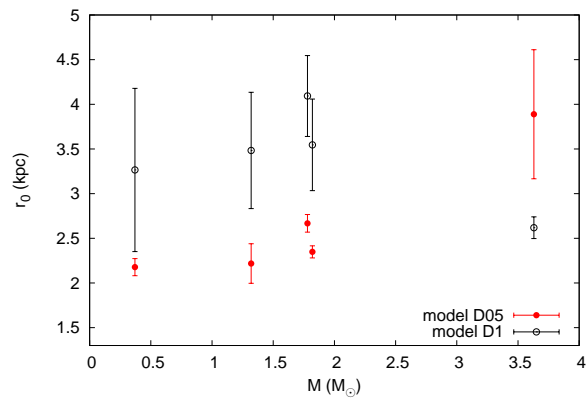}
\caption{Initial positions of the GCs in model D05 (filled red circles) and model D1 (open black circles). The error bars are obtained assuming a $25\%$ error on the evaluation of the orbital eccentricity.}
\label{mapICD}
\end{figure}

Figure \ref{mapICD} shows the initial positions of our GCs obtained in this way for each of the 2  models considered. The error bars are obtained assuming an error over the eccentricity of $\sim 25\%$. These semi-analytical results are strong hints in favour of an in-situ origin for all the clusters in both models, because all the clusters have $r_0 < R_t$, where $R_t=5$ kpc.

\begin{table}
\caption{}
\centering{Main observed properties of the Fornax GCs}
\begin{center}
\begin{tabular}{lcccccc}
\hline
\hline
  ID  & $M_{\rm GC}$ & $R_{\rm GC}$ & $v_{\rm GC}$ & $r_{\rm GC,c}$ \\
	& $(10^{5}$ M$_\odot)$ & (kpc) & (km s$^{-1}$) & (pc) \\
\hline
GC1 & 0.37 & 1.60 & -             & 5.01 \\
GC2 & 1.82 & 1.05 & $-1.2\pm 4.6$ & 5.81 \\
GC3 & 3.63 & 0.43 & $ 7.1\pm 3.9$ & 1.60 \\
GC4 & 1.32 & 0.24 & $ 5.9\pm 3.4$ & 1.75 \\
GC5 & 1.78 & 1.43 & $ 8.7\pm 3.6$ & 1.38 \\
\hline
\end{tabular}
\end{center}
\begin{tablenotes}
\item Column 1: GC name. Column 2: mass of the GC. Column 3: projected distance to the Fornax centre. Column 4: line-of-sight velocity with respect the Fornax centre of velocity. Column 5: GC core radius.
\end{tablenotes}
\label{GCobs}
\end{table}

We carried out our simulations up to $\sim 3$ Gyr for both the models discussed below. 

In model D05, this time is twice the time needed for the decay of the heaviest GCs, which are the only contributing to the formation of a nucleus.

On the other hand, in model D1, the GCs orbital infall and disruption is much faster, occurring in $\lesssim 1$ Gyr. In this case, we carried out the simulation up to $3$ Gyr in order to investigate whether the GCs remnant can slowly deposit around the Fornax centre and give rise to a detectable nucleus.

\subsection{Model D05}

\begin{table*}
\caption{}
\centering{Main parameters of GCs in model D05}
\begin{center}
\begin{tabular}{lcccrccrcrr}
\hline
\hline
   \multicolumn{1}{c}{ID}  & $W0$ & $M_{\rm GC}$ & $r_{\rm GC}$ & \multicolumn{1}{c}{$r_{\rm GC,t}$} & $\sigma$ & ${\rm Log} ~\rho_0$ & \multicolumn{1}{c}{N} & e & \multicolumn{1}{c}{$\tau_{\rm df}$} & \multicolumn{1}{c}{$\tau_{\rm td}$}\\

	& &$(10^{5}$ M$_\odot)$ & (kpc) & (pc) & \multicolumn{1}{c}{(km s$^{-1}$)} & $($M$_\odot$ pc$^{-3})$ &  & &\multicolumn{1}{c}{(Gyr)} &\multicolumn{1}{c}{(Gyr)} \\
\hline
\multicolumn{11}{c}{model D05}\\
\hline
GC1 & $4.9$ &$ 0.37$ & $ 1.60$ & $102$ & $ 1.25$ & $ 0.41$ & $ 3346 $  &$0.6$ &$21.4$&$7$\\
GC2 & $7.1$ &$ 1.82$ & $ 1.05$ & $200$ & $ 1.92$ & $ 1.26$ & $16463 $  &$0.3$ &$16.3$&$20$\\
GC3 & $7.5$ &$ 3.63$ & $ 0.44$ & $ 82$ & $ 4.36$ & $ 3.09$ & $32837 $  &$0.8$ &$1.6$&$2.0$\\
GC4 & $6.7$ &$ 1.32$ & $ 0.42$ & $ 50$ & $ 3.39$ & $ 2.79$ & $11940 $  &$0.6$ &$0.6$&$2.0$\\
GC5 & $8.6$ &$ 1.78$ & $ 1.45$ & $151$ & $ 2.26$ & $ 2.65$ & $16101 $  &$0.4$ &$6.9$&$12.0$\\
\hline
\multicolumn{11}{c}{model D1}\\
\hline
GC1 & $4.6$ &$ 0.37$ & $ 1.61$ & $  91$ & $ 1.32$ & $ 1.46$ & $3376 $   &$0.9$ &$4.4$&$0.3$\\
GC2 & $6.4$ &$ 1.82$ & $ 1.48$ & $ 144$ & $ 2.33$ & $ 1.42$ & $16606 $  &$0.7$ &$2.8$&$0.5$\\
GC3 & $7.5$ &$ 3.63$ & $ 0.55$ & $  83$ & $ 4.33$ & $ 3.08$ & $33121 $  &$0.0$ &$0.6$&$0.7$\\
GC4 & $6.3$ &$ 1.32$ & $ 0.54$ & $  38$ & $ 3.85$ & $ 2.90$ & $12044 $  &$1.0$ &$0.2$&$3.2$\\
GC5 & $9.3$ &$ 1.78$ & $ 2.42$ & $ 220$ & $ 1.86$ & $ 2.48$ & $16241 $  &$0.7$ &$19.1$&$2.8$\\
\hline
\end{tabular}
\end{center}
\begin{tablenotes} 
\item Column 1: name of the GC. Column 2: adimensional potential well. Column 3: total mass. Column 4: radial distance to the Fornax centre. Column 5: tidal radius. Column 6: velocity dispersion. Column 7: central density. Column 8: number of particle used to model the GC. Column 9: orbital eccentricity. Column 10: df time-scale evaluated through Equation \ref{tdf}. Column 11: tidal disruption time-scale as extrapolated from Figure \ref{traj05}.
\end{tablenotes}
\label{GCnbo}
\end{table*}

In this case, the Fornax density profile is significantly shallower than what expected from standard CDM predictions.

\begin{figure*}
\centering
\includegraphics[width=16cm]{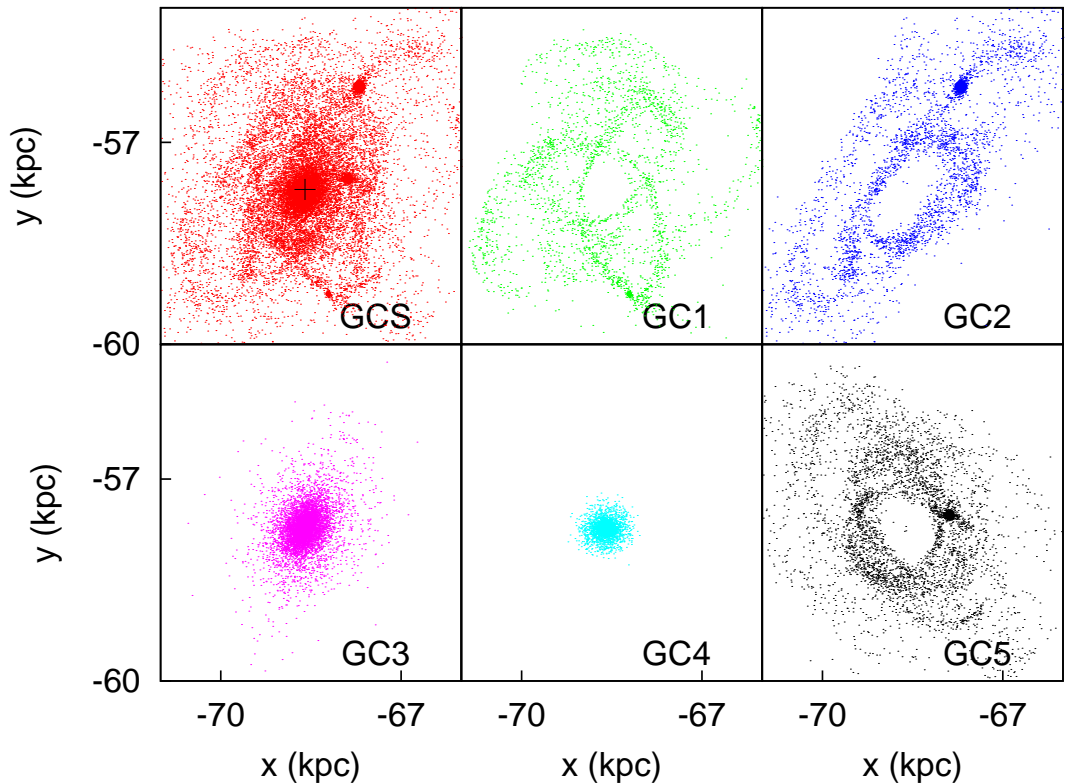}
\caption{The Fornax GCS after $3.1$ Gyr in model D05 in the MW reference frame. In the top panel on the left column is represented the whole GCS, whereas from top to bottom and from left to right in each one of the other panels are represented the five GCs. 
The white filled circle in the left corner top panel identifies the Fornax centre of density. }
\label{GCS05}
\end{figure*}

In order to highlight the effect of tidal forces and dynamical friction of the GC motion, we show in Figure \ref{GCS05} a snapshot at $3.1$ Gyrs of the various GCs. Looking at the figureit is evident that tidal forces have significantly acted on cluster GC1, GC2 and GC5. However, while GC2 and GC3 keep a clearly visible core, GC1 is almost completely dispersed along its orbit. On the other hand, clusters GC3 and GC4 have been able to reach the centre of Fornax, thus contributing to the formation of a compact structure. Note that in this case the coordinates of the Fornax center in the MW reference frame are $\mathbf{-68.5,-58,-161}$ kpc in the MW reference frame, as highlighted in the figure.
 
 If we look at the GCS as a whole, which is shown in the top left panel of the figure, we find that GCs merging have produced a quite evident central structure, which is orbited by 3 small clumps, relics of GC1, GC2 and GC5.

\begin{figure}
\centering
\includegraphics[width=8cm]{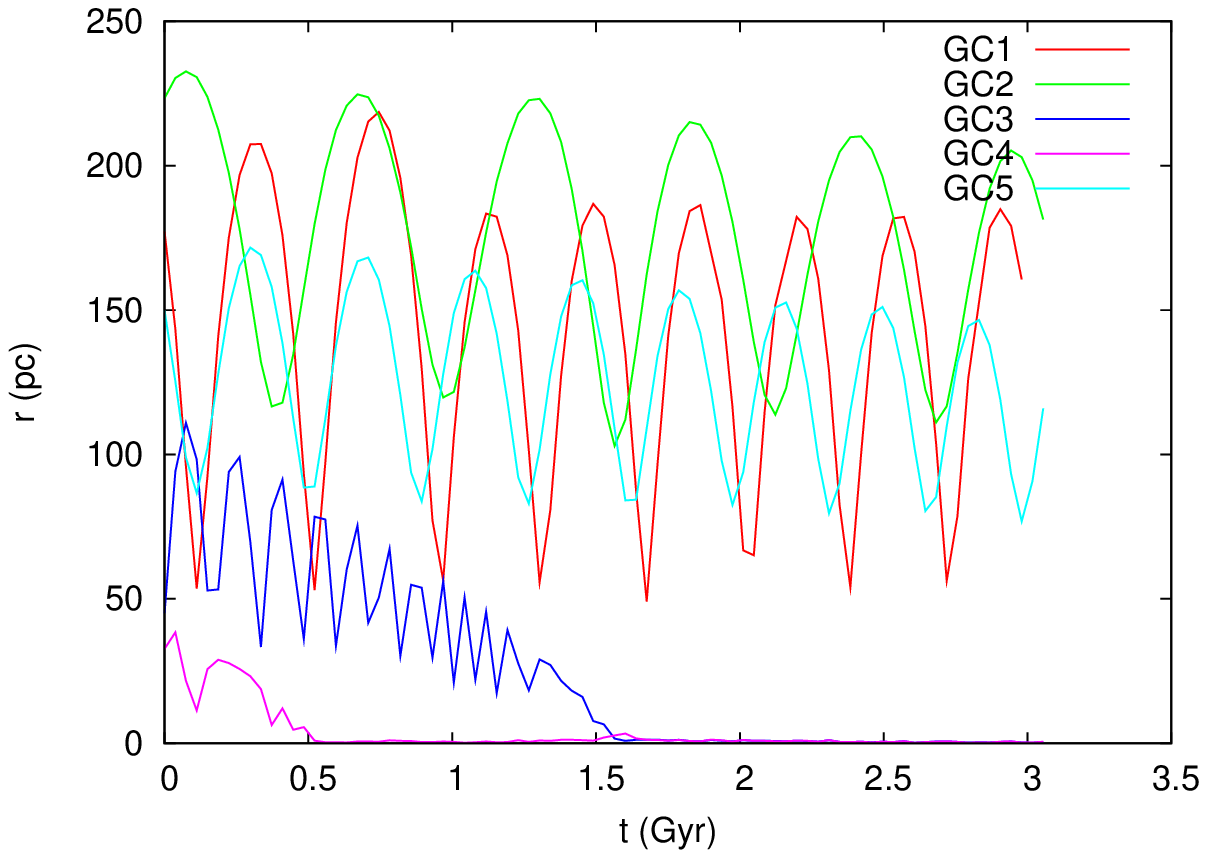}\\
\includegraphics[width=8cm]{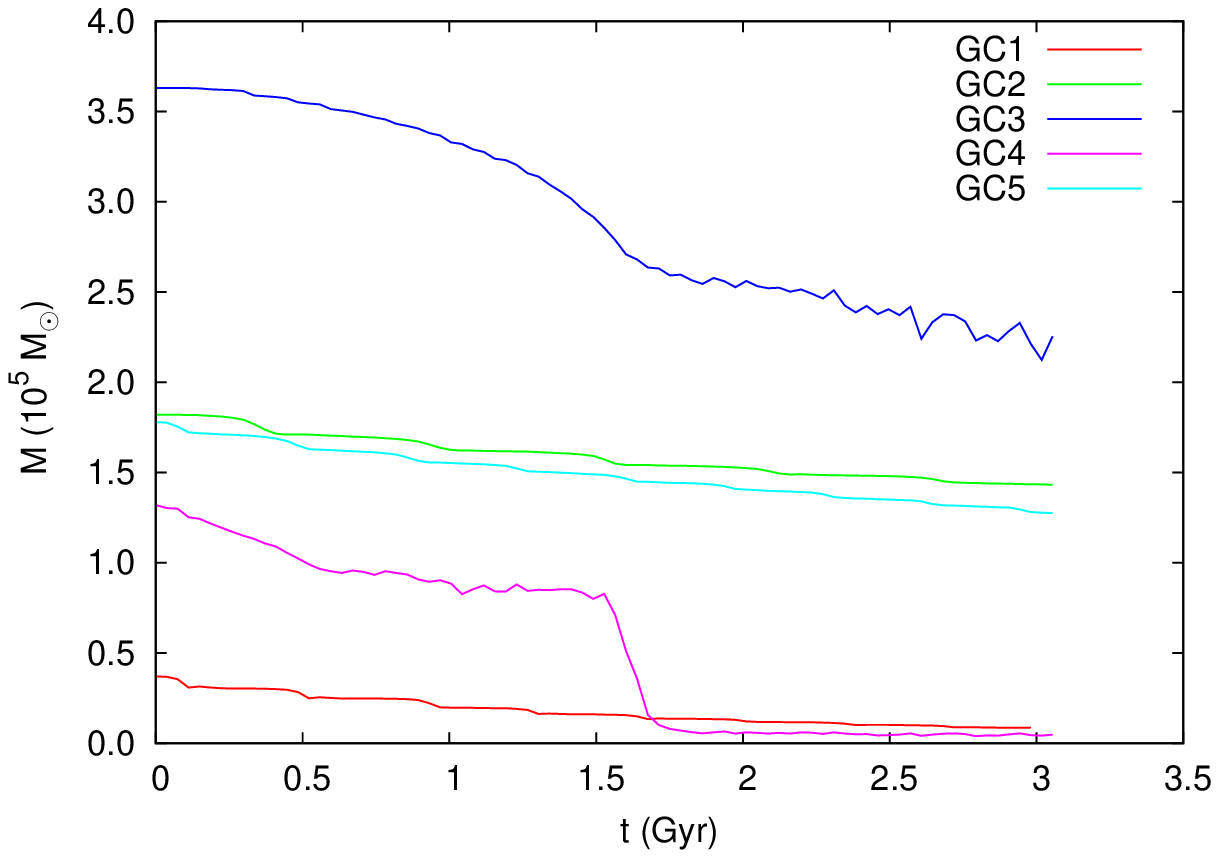}\\
\caption{Time evolution of the radial distance to the Fornax centre (top panel) and of the total mass (bottom panel) for all the GCs in model D05.}
\label{traj05}
\end{figure}

The upper panel in Figure \ref{traj05} shows the evolution of the GCs radial 
their mass as a function of time. To estimate the GC mass, we evaluate at each time-step its tidal radius using Eq. \ref{tidR}, conveniently adapted to the GC motion within Fornax. Then, we assume that the GC mass is that enclosed within such radius.

The clusters GC3 and GC4 merge into the galactic centre within $1.5$ Gyr, dragging there $\sim 2.6 \times 10^5 $ and $\sim 0.5\times 10^5$ M$_\odot$, respectively.

On the other side,  the bottom panel of Fig. \ref{traj05} allows to extrapolate the time at which the GC mass becomes smaller than $10^4$ M$_\odot$, which we take as an estimate of the time required by the tidal action of the background stars to disrupt the incoming clusters ($\tau_{\rm td}$).
In particular, we find $\tau_{\rm td}<\tau_{\rm df}$ for clusters GC1, GC2 and GC5 that means they poorly contribute to the formation of a NC.

\begin{figure}
\centering
\includegraphics[width=8cm]{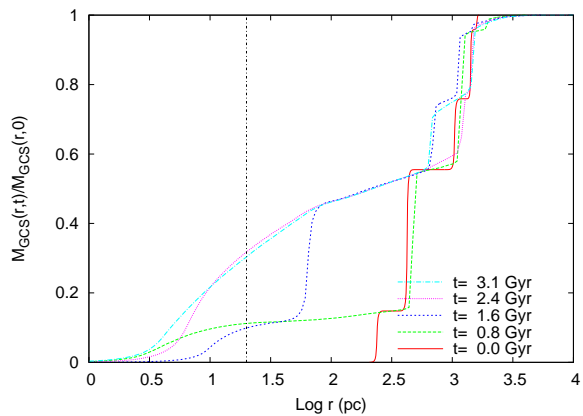}
\caption{Cumulative, mass distribution of the GCs at different times in model D05.}
\label{Mr05}
\end{figure}

Figure \ref{Mr05} shows the mass distribution of the 5 GCs at different times; the mass enclosed within $\sim 20$ pc saturates after $2$ Gyr to a value that is $\sim 30\%$ of their total mass, i.e. $\simeq 2.6\times 10^5$ M$_\odot$. 

\begin{figure}
\centering
\includegraphics[width=8cm]{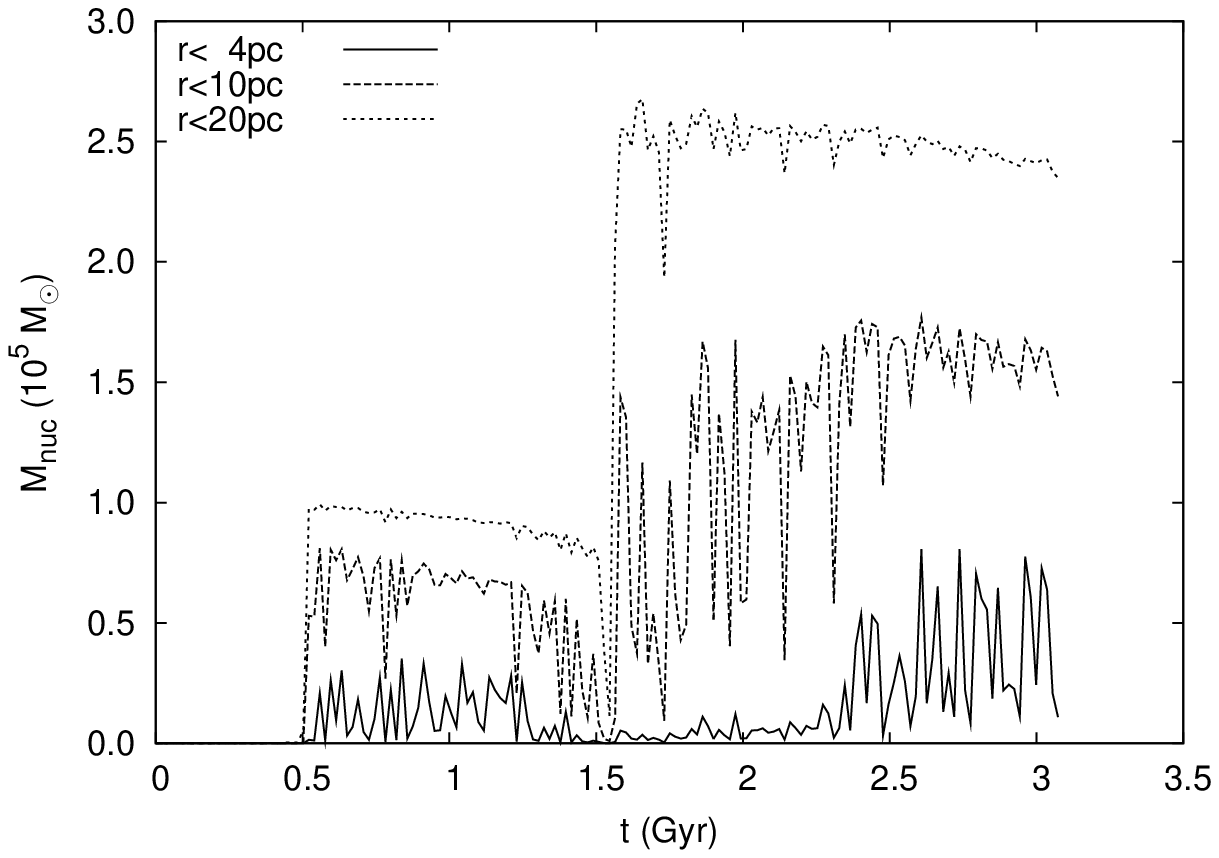}
\caption{Mass deposited within 4 pc (straight line), 10 pc (dashed line) and 20 pc (dotted line) from the Fornax centre as a function of the time in model D05.}
\label{NSC05}
\end{figure}

Some hints about the formation of a NC can be obtained looking at the amount of mass deposited into the galactic centre. For instance Fig. \ref{NSC05} shows the mass enclosed within 4, 10 and 20 pc from the Fornax centre as a function of time. It is evident a steep increase that corresponds to the  orbital decay of clusters GC2 and GC3.

\begin{figure}
\centering
\includegraphics[width=8cm]{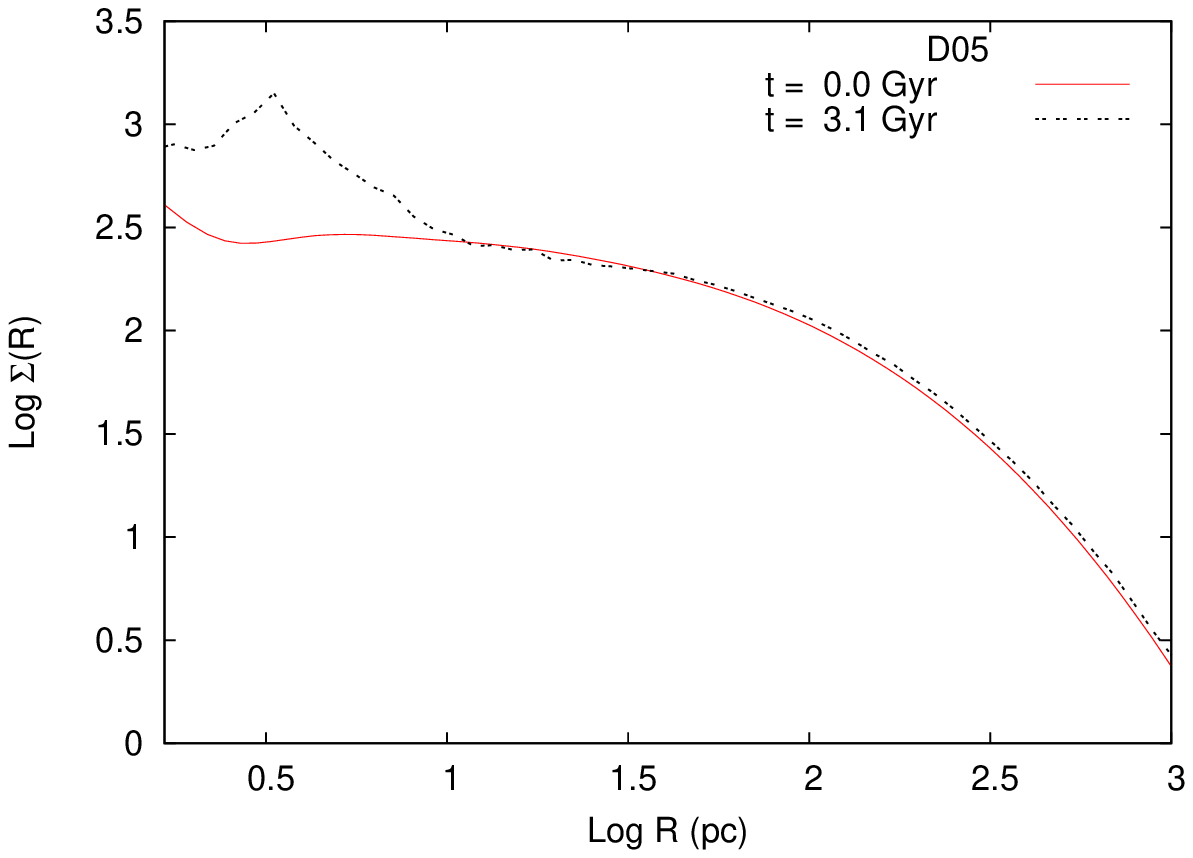}
\caption{Surface density profile of the Fornax dSph at the beginning of the simulation and after $3.2$ Gyr in model D05. It is evident a nucleus with effective radius $r_{NC}\simeq 5.6$ pc.}
\label{Sigma05}
\end{figure}

To check whether the accumulated mass in model D05 leads to the formation of a detectable nucleus, we show in Figure \ref{Sigma05} its surface density profile. The formation of a ``bright'' nucleus characterised by an effective radius $r_{\rm NC} \simeq (5.6\pm 0.8)$ pc is evident. The enclosed mass is $M_{\rm NC} \simeq (1.03\pm 0.07) \times 10^5$ M$_\odot$. 

\subsection{Model D1}

In the more centrally concentrated model D1, tidal forces acting on GCs are much more effective.

\begin{figure*}
\centering
\includegraphics[width=16cm]{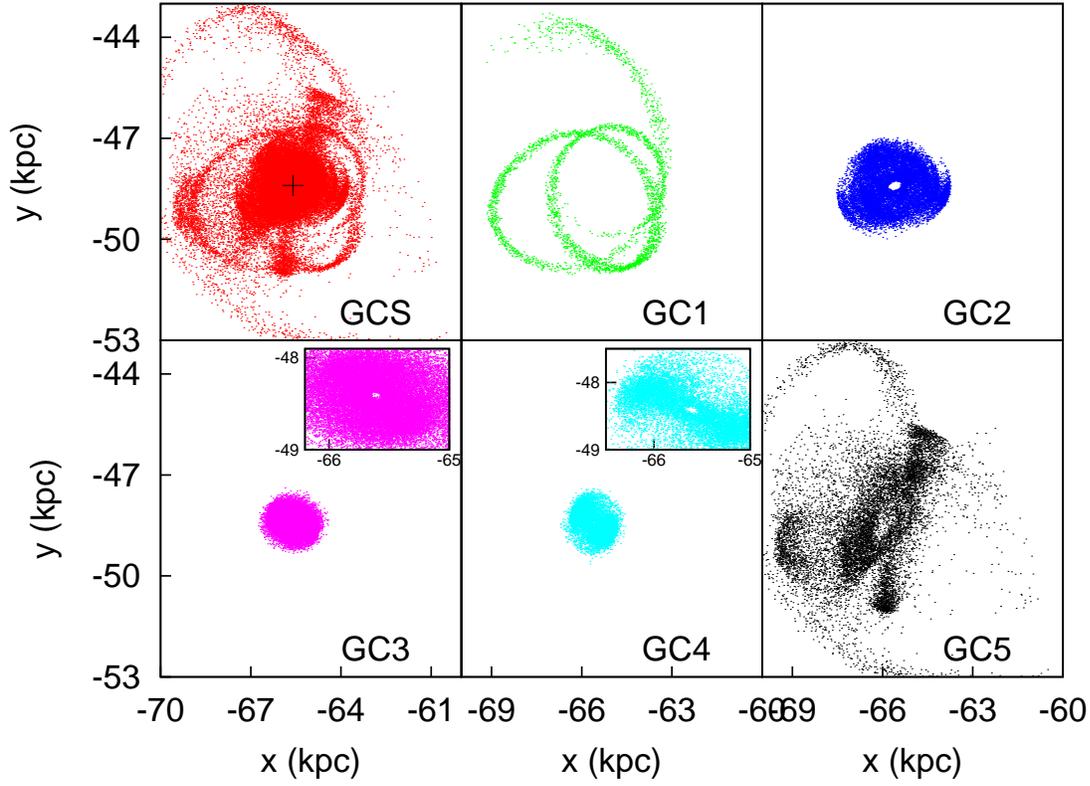}
\caption{As in Figure \ref{GCS05}, but for model D1. In this case, looking at the zoomed panels contained in the bottom row, it is evident that GC3 and GC4 cannot penetrate the innermost region of Fornax. Indeed in this case there is a hole in their centre.}
\label{GCS1}
\end{figure*}

Figure \ref{GCS1} shows the whole GCS after $3.1$ Gyr and also each cluster, making clear which of them contribute more to the formation of a NC. 
In this case, GC1 and GC5 are clearly disrupted by tidal forces, whereas GC2, GC3 and GC4 resist during their decay to the Fornax centre.
In particular, it is worth noting that most of the mass lost by GC5 is deposited, after $3.1$ Gyr, into the Fornax tidal tail, contributing significantly to its enhancement. 

\begin{figure}
\centering
\includegraphics[width=8cm]{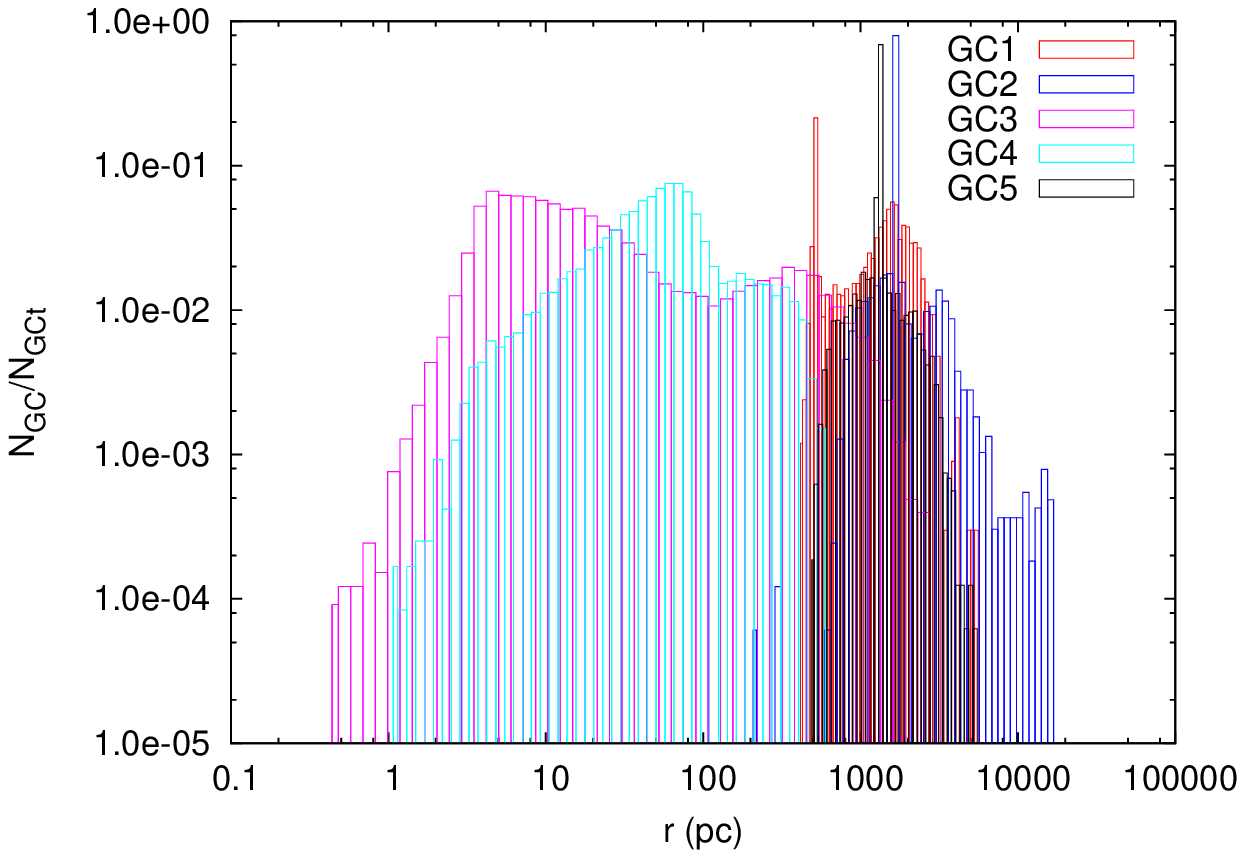}\\
\includegraphics[width=8cm]{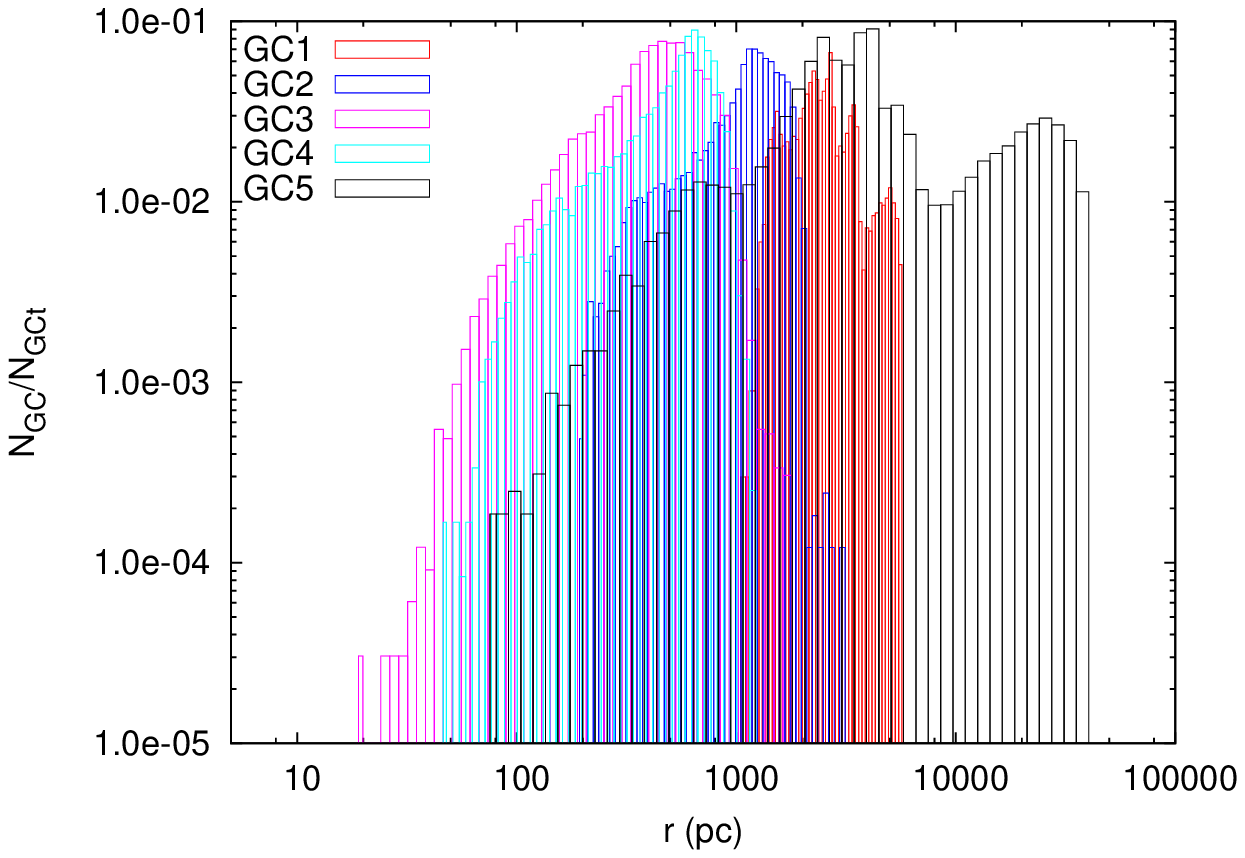}\\
\caption{Fraction to the total of the number particles in each GC, as a function of their distance to the Fornax centre. Top panel: model D05. Bottom panel: model D1.}
\label{hysto}
\end{figure}

In Figure \ref{hysto}, we compare the number distribution of the GCs particles as a function of their distances from the Fornax centre, for both model D05 and D1. It is evident that in model D1 the GC5 cluster is completely disrupted, with some of its debris dispersed on distances beyond $15$ kpc. 

Surprisingly, we found that none of the clusters can deposit mass in the innermost region of Fornax ($r\lesssim 50$ pc). Indeed, the GCs debris seem to distribute around a sort of ``hole''. 

Actually, this region is not really empty, since it contains background stars. Actually, it seems that the high density of the background prevents the deposit of mass in the innermost region of the galaxy. 
Hence, the background steep density profile acts as a MBH does in bright galaxies, suppressing the formation of a nucleus.

\begin{figure}
\centering
\includegraphics[width=8cm]{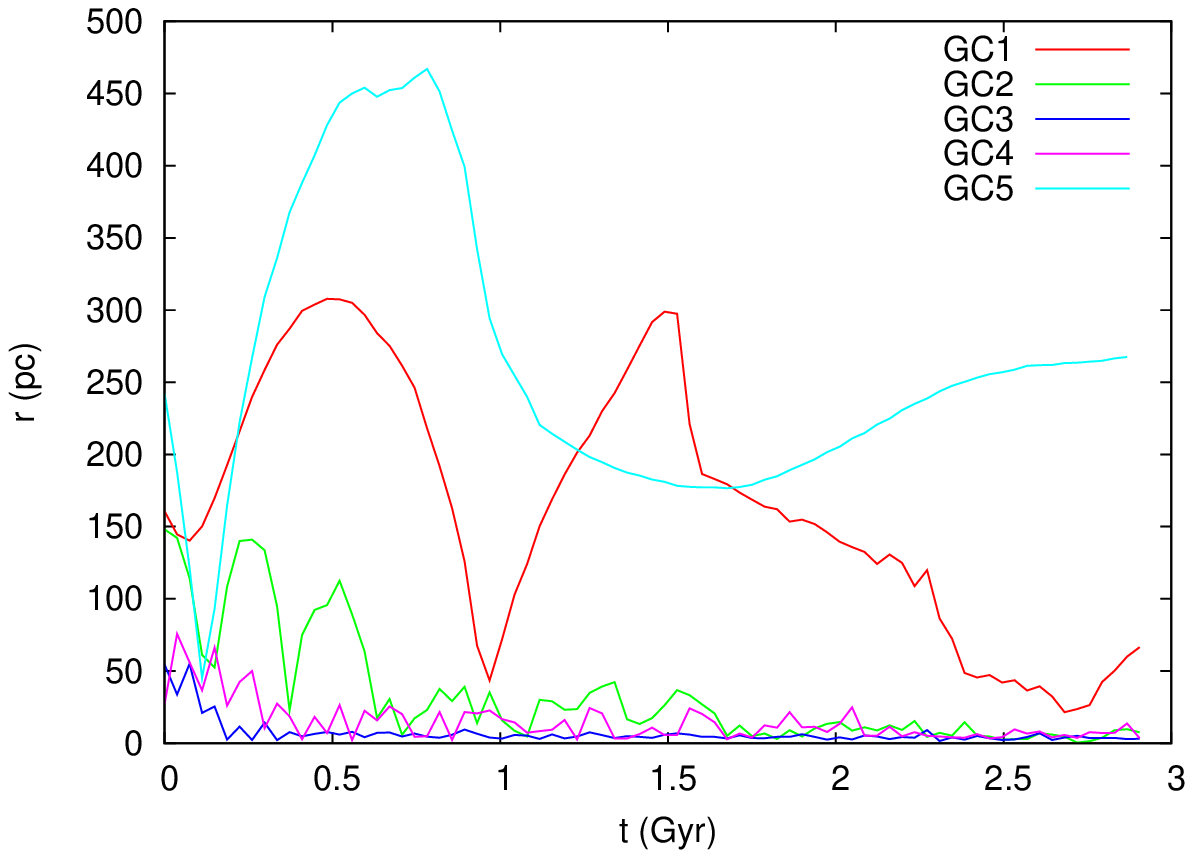}\\
\includegraphics[width=8cm]{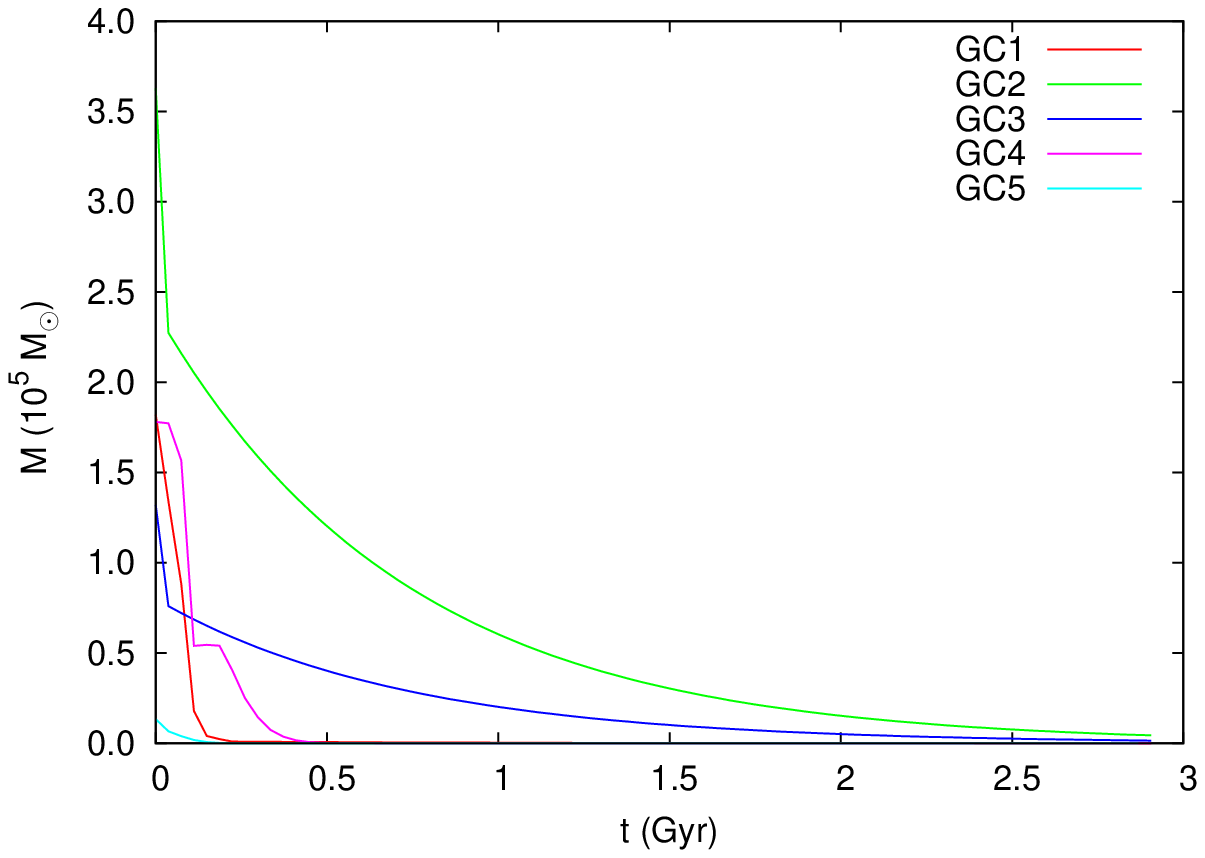}\\
\caption{As in Figure \ref{traj05}, but for model D1.}
\label{traj1}
\end{figure}

Indeed, a comparative look at the time evolution of the GC radial positions and masses (Fig. \ref{traj1}) shows that GC1 and GC2 are disrupted within 0.3 Gyr, well before they reach the inner 100 pc of the Fornax. Also GC5 is almost completely disrupted from tidal forces in 0.5 Gyr. Clusters GC3 and GC4, instead, are able to reach the innermost region of the galaxy, though they suffer a loss of more than $80\%$ of their initial masses.

\begin{figure}
\centering
\includegraphics[width=8cm]{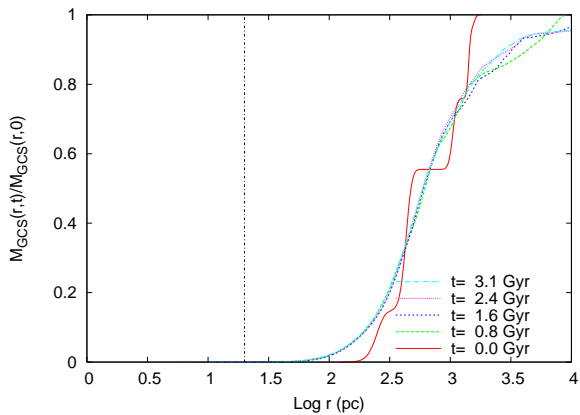}
\caption{As in Figure \ref{Mr05}, but for model D1.}
\label{Mr1}
\end{figure}

\begin{figure}
\centering
\includegraphics[width=8cm]{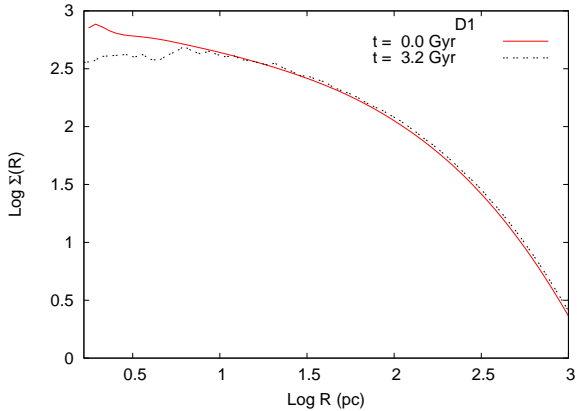}
\caption{As in Figure \ref{Sigma05}, but for model D1.}
\label{Sigma1}
\end{figure}

The efficiency of tidal forces is well evident in Figure \ref{Mr1}, which shows the cumulative mass distribution of the GCs at different times. 
Figure \ref{Sigma1} shows, indeed, that the orbital evolution of the GCs in this case do not lead to the formation of a detectable nucleus. 

\section{Consequences for MBH formation}

\begin{figure}
\centering
\includegraphics[width=8cm]{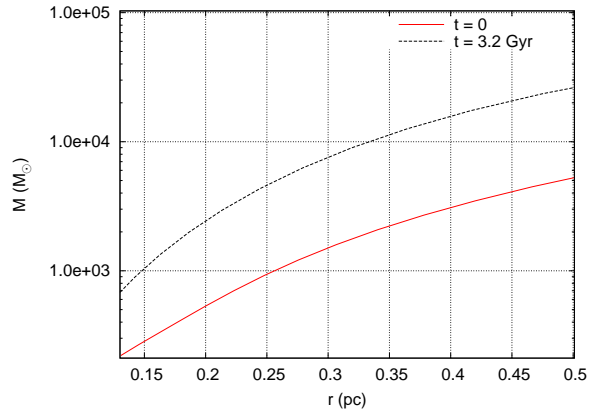}
\caption{Enclosed mass as a function of radius for Fornax at the beginning of the simulation and after $3.1$ Gyr for model D05.}
\label{D05mas}
\end{figure}

In model D05, the GC orbital decay gives rise to a well evident nucleus, whose surface density profile 
exceeds $\sim 5$ times the background density. In particular, we show in Figure \ref{D05mas} the cumulative radial mass distribution at $t=0$ and after $3.1$ Gyr for the whole system (Fornax+GCS). It is evident that the mass enclosed within $0.1$ pc exceeds $M_{\rm enc}=10^3$ M$_\odot$, corresponding to a central density of $\sim 2\times 10^5$ M$_\odot$ pc$^{-3}$. This value is about a hundred times higher than the initial central density of the most massive cluster, GC3, which is the one that contributes mostly to the NC formation.

As known, during their evolution GCs undergo mass segregation, which can lead to the formation of massive sub-systems in their cores, mainly composed by stellar BHs (see for example \cite{spitzer69,BAPZ03,BAUMEBIS,as16} and reference therein). 
The subsequent evolution of such systems depend on the cluster properties and may lead, in some cases, to the formation of a massive black hole seed. This formation process has been usually invoked to explain the formation of the so-called intermediate mass black holes (IMBH), with masses ranging between $10^2-10^4$ M$_\odot$, in dense and massive star clusters \citep{zwart02,freitag06a,gaburov08,mackey08,giersz15,as16}.

In particular, for metal poor systems, \cite{as16} has shown that mass segregation drives the formation of a massive central system of evolved stars composed for $\sim 60\%$ by stellar BHs. Dynamical interactions among these BHs possibly leads to the formation of an IMBH seed upon certain conditions \citep{giersz15} but, in other cases, they can survive within the GC centre up to a Hubble time \citep{wang16}.

In any case, the formation of a dense sub-cluster of dark objects or an IMBH seed as a consequence of mass segregation and strong dynamical interactions is a process with typical time scales that increase at increasing the mass of the host system. Indeed, this process occurs over several times the relaxation time-scale, which, approximately, scales with the host mass. For instance, for a typical galactic nucleus, the time-scale for such process easily exceeds a Hubble time. However, it is considerably smaller for smaller star clusters.

In this framework, our results can be used to investigate whether the orbital decay can lead to the formation, within the Fornax centre, of a massive black hole seed. 
As shown in Figure \ref{GCS05}, the formation of a NC is mainly due to the orbital decay of clusters GC3 and GC4. The scaling relation provided by \cite{as16}, suggests that these two metal poor clusters should host at their centre two massive systems mainly composed of heavy stellar remnants, with masses of $2090\pm 48$ M$_\odot$ and $758\pm 18$ M$_\odot$, respectively, and sizes of $\sim 0.1$ pc. 

In their current version, our simulations cannot be used to follow stellar collisions, then we cannot state anything about the possible formation process of an IMBH. However, we can guess something about the possible accumulation of stellar BHs in the Fornax centre. Indeed, on a side we know that mass segregation will lead to a significant deposit of heavy stellar BHs in the centre of the infalling GCs, while on the other side we know from Figure \ref{hysto} that the particles that are deposited into the Fornax centre are those moving in the innermost region of the host cluster. Therefore, we expect that during the GCs merging their cores, composed mainly of heavy stellar remnants, will collide together.
The subsequent increase in the surrounding density can turn on a phase of strong dynamical interactions, that can rapidly lead to the formation of a seed, as suggested by \cite{giersz15}. 

On the other hand, if at least one of the infalling GCs hosts a IMBH, which has formed during its past evolution, it should settle at the Fornax center due to dynamical friction, and subsequently may undergo a slow accretion phase. 

It should be noted that if the IMBH is transported to the galactic centre by the infalling cluster, its mass will not scale with the host galaxy mass as expected for super-massive black holes (SMBHs), since in this case the central black hole would be significantly lighter. Indeed, for a Fornax-like system the expected mass for a central SMBH would be $M_{\rm BH} = 1.5\times 10^4$ M$_\odot$, according to the scaling relation provided by \cite{scot}, $10$ times heavier the mass of the infalling GC IMBH. Of course, such small IMBH would have poor effects on the surrounding nucleus, making its identification even more difficult.

Hence, if the dSph is characterised by a flat core or a shallow cusp at its formation, the decay and merging of its GCs can transport a significant amount of matter toward the galactic centre, causing an increase in the central density that can potentially favour the formation, or accretion, of a massive black hole. 
On the other hand, it should be noted that among all the dwarf spheroidal galaxies in the local group, only the Sagittarius dSph likely displays a nucleus in its centre, despite its origin is still poorly understood. 
In this framework, hence, the absence of bright nuclei in dSphs would imply the absence of central massive black holes but, at the same time, the possible future formation of such nuclei can facilitate the formation of a central IMBH.

\section{Implications for the core/cusp problem}
\label{discussion}

In this paper we showed that the density profile of the Fornax dSph plays a fundamental role in determining the fate of its GCs. Indeed, we found that if Fornax is characterised by a steep density profile, its GCs will likely be disrupted in $\lesssim 1$ Gyr. On the other hand, a shallower density profile would allow the formation of a relatively small NC, with a mass few times $10^5$ M$_\odot$.

This allows some inference on the Fornax GCS history. 

Indeed, let's make the, likely, hypothesis that the galaxy hosted, in the past, a larger population of clusters. The additional GCs should had smaller apocentres than those currently observed, since they should have completely accomplished their orbital decay. Anyway, our results suggest that, even if these GCs existed in the past, the efficient tidal action of the stellar background could have completely dispersed their debris, hiding the presence of any signature of their existence. 
This possibility represents an interesting alternative explanation for the the so-called Fornax ``\textit{timing problem}'', which states that is extremely unlikely that we are looking at the GCs just before their final sink to the Fornax centre. Indeed, it would imply that a steep stellar cusp in the galaxy centre can halt the formation of a nucleus, disrupting efficiently the infalling clusters and covering their past existence in the galactic background.

These results indicate that a galactic nucleus acts in the same way as SMBHs do, shattering the infalling GCs and halting the formation of a central nucleus \citep{Dolc93,Antonini13,ascd15lack}.
The effect of GC disruption operated by a steep cusp in the host galaxy density profile may be responsible for the observational absence of NCs in small dwarf galaxies (with masses around $10^8$ M$_\odot$). 

Interestingly, the absence of a nucleated region can tell us something about the luminous and dark matter content of the smallest, DM dominated, galaxies. Indeed, it is widely believed that dSph galaxies are characterised by large quantities of DM, but with density profiles significantly shallower than what expected by standard CDM paradigm. This is known as core/cusp problem. One of the most credited scenario propose that initially a DM halo form with a cuspy density profile, which is subsequently removed by merging events, tidal interaction with a larger galaxy, stellar formation, gas accretion and supernovae (SN) explosions (see for example \cite{mayer01, yozin, pontzen12, governato12, teyssier13, kazantzidis13, nipoti15,pontzen15} and reference therein). 
All these mechanisms can remove the initial DM cusp over time-scales of the order of few Gyr, leaving the dSph with a nearly cored density profile, or, at least, a mild cusp ($\gamma\leq 0.6$).

In a recent work, \cite{read16} have shown that SN feedback and star formation drives the formation of a core on a time $\sim 4$ Gyr, a time which is fully compatible with the time-scale over which GCs orbital decay can occur. This may imply that 
cusp/core transformation is a complex process in which several different processes act together with an efficiency that likely depends on the host galaxy properties. 

\begin{figure}
\centering
\includegraphics[width=8cm]{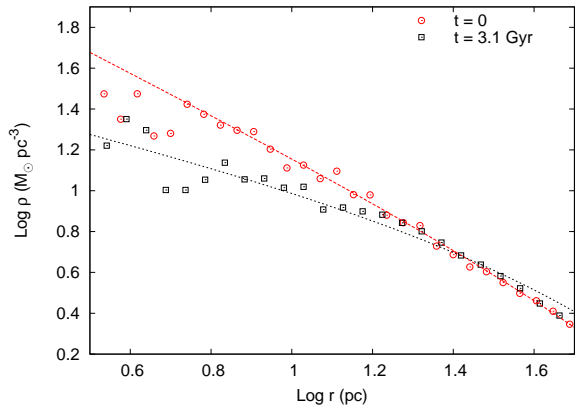}
\caption{
Density profile of the Fornax galaxy and its GCs at $t=0$ and $t=3.1$ Gyr in model D1. The initial cusp significantly flattens by the end of the simulation.}
\label{D1rho}
\end{figure}

Moreover, another outcome of our results that can be related to the cusp/core problem is the following. Looking at model D05, we demonstrated that the future orbital decay of GCs will likely drive the formation of a NC, well visible as an edge in the surface density distribution. However, no clear signature of an NC is found in model D1. Indeed, looking closer to the Fornax centre, we found that its spatial density profile seems to flatten, as shown in Figure \ref{D1rho}. Surprisingly, a best-fit of the density profile with a Dehnen model reveals two important modifications of the profile: i) the scale radius decreases, passing from $391$ to $285\pm 12$ pc and ii) the inner slope $\gamma$ significantly decreases, passing from $1$ to $0.65\pm 0.04$. Therefore, the early orbital decay of GCs in DM dominated dSph seems to be a complementary explanation for the core/cusp problem. This possible solution to the problem seems  very general, as observations indicate that GC formation is an ubiquitous process occurring in all the galaxies. This alternative process for the cusp removal was firstly proposed by \cite{Goerdt08} and \cite{Goerdt10}. In particular, \cite{Goerdt10} demonstrated that very massive, point-like satellites can remove the cusp in DM halos at very low densities ($\lesssim 10^{-2} $ M$_\odot$ pc$^{-3}$).

Despite the mechanism observed in our simulations is  analogous to the one discovered by \cite{Goerdt10}, the crucial difference is that in this case it acts on relatively smaller scales, shaping the properties of dSph-sized systems. Moreover, our results suggest that cusp/core transition can be operated by the orbital decay of ``normal'' GCs, with masses $\lesssim 10^5$ M$_\odot$.
More in detail, the models presented here are different, for three main reasons: 

\begin{itemize}
\item we modelled each GC using a reliable mass function as suggested by observations;
\item we modelled each GC by a sample of particles, thus accounting for their internal dynamics and their response to external tidal forces;
\item we modelled the evolution of relatively light GCs, with masses below $5\times 10^6$ M$_\odot$, traversing a cuspy galaxy with typical density $\rho > 100$ M$_\odot$ pc$^{-3}$.
\end{itemize}
Even in such different conditions our results show that the sinking GCs can remove efficiently the cusp of a DM dominated dSph.

\section{Discussion and conclusions}
\label{conclusion}
 
In this paper we investigated the future evolution of the five GCs observed in the Fornax dSph. Our results suggest that the GCs orbital decay can lead to a detectable NC only if the galaxy has a density profile shallower than $r^{-1}$. Otherwise, the galactic density cusp acts as a tidal heater, efficiently disrupting the incoming clusters before they can reach the galactic center and deposit mass therein to give rise to a detectable nucleus. 

Extending our results to a generic dSph that host some GCs close to its centre, we have shown that the GC infall can facilitate the accumulation
of stellar BHs within the galactic central region, as well as of an IMBH formed during the past GC evolution. This would have implications on the possible BH-galaxy mass scaling relation, since for dSph the central black holes would be significantly lighter than expected, and hardly detectable through usual ``dynamical estimators'', such as the projected velocity dispersion \citep{vandermarel10,haggard13}.
Moreover, we have shown that if the dSph has a steep cusp at its birth, as expected by cosmological simulations, clusters traversing the galactic region are efficiently disrupted by tidal forces, thus quenching the formation of an NC. This would explain the absence of bright nuclei in the dSphs in the local group. Indeed, one possible path to the dSph formation and evolution is the following:
\begin{itemize}
\item[I)] the dSph forms with a steep cusp in the density profile and some GCs form following the overall galactic distribution;
\item[II)] due to dynamical friction and tidal forces, loose GCs moving on inner orbits are efficiently disrupted, leaving almost untouched only those moving out of the galaxy scale radius;
\item[III)] the cusp distribution transforms into a cored one, as a consequence of some process, as explained in Section \ref{discussion};
\item[IV)] the surviving clusters undergo dynamical friction and, due to the reduced efficiency of tidal forces, eventually reach the galactic centre and drive the formation of a bright nucleus.
\end{itemize}

In this picture, the GCs tidal disruption could represent an efficient process for transforming the cuspy profile into a flatter one, as shown in the previous section. 

In the following, we briefly summarize the main outcomes of this work: 
\begin{itemize}
\item using highly-resolved, direct, $N$-body simulations, based on observations of the Fornax dSph, we modelled the evolution of a globular cluster system within the Fornax dwarf spheroidal galaxy, representing in a reliable way both the clusters and the parent galaxy;
\item we have shown that if the dSph has a shallow cusp in the density profile, the GC orbital decay leads to the formation of a detectable nucleus over a time-scale of $1.5$ Gyr, with properties similar to those expected by extrapolation of observational scaling relations valid for heavier galaxies;
\item on the other hand, if the dSph has a cuspy density profile, as suggested by CDM predictions, tidal forces are strong enough to disrupt the infalling clusters, thus preventing the formation of a NC;
\item in the case of nearly cored dSphs, we expect that GC decay and merging can favour the formation of a IMBH, especially if one of the GCs contains an IMBH previously formed, this would imply a BH-galaxy mass relation substantially steeper at low galaxy mass, with BH masses significantly smaller than expected from observational scaling relations based on heavier galaxies data;
\item in the case of cuspy dSphs, instead, we showed that the GCs tidal disruption significantly affects the background mass distribution. In particular, the density profile significantly flattens. This process can play an interesting role in the solution of the core/cusp problem;
\item the missing observational evidence of nuclei in the LG dSphs would imply that dSphs have formed with a steep cusp, thus disrupting all those clusters formed within the galactic scale radius; after the transformation of the cusp into a core, only the surviving clusters can lead to the formation of a nucleus, but on a significantly longer time-scale, since they formed in an outer region of the galaxy;
\item finally, we suggest that dSph may follow a well determined evolutionary path: i) they form with a steep density profile, ii) the innermost star clusters are disrupted by tidal forces and, in turn, they flattens the galactic density profile, iii) at this point, farther GCs can reach the galactic centre and iv) contribute to the formation of a nucleus.
\item The clear dearth of bright nuclei in the LG dSph population may indicate that the first part of the evolutionary process (points i) and ii)) has a very long time, as also suggested by simulations.
\end{itemize}

\section{Acknowledgement}

MAS acknowledges financial support from the University of Rome ``Sapienza'' through the grant ``52/2015'' in the framework of the research project ``MEGaN: modelling the environment of galactic nuclei''.
Part of this work was performed at the Aspen Center for Physics, which is supported by National Science Foundation grant PHY-1066293. At this regard, RCD thanks the Simons foundation for the grant which allowed him a period at the Aspen Center for Physics where he developed part of this work.

\clearpage
\footnotesize{
\bibliographystyle{mn2e}
\bibliography{bblgrphy2}
}

\end{document}